\documentclass{aa}  

\usepackage{graphicx}
\usepackage{txfonts}
\usepackage{xcolor}
\usepackage{hyperref}
\usepackage{listings}
\usepackage{comment}
\begin{document}

   \title{The effects on structure of a momentum coupling between dark matter and quintessence}

   \author{G. N. Candlish
          \inst{1}
          \and
          Y. Jaffé\inst{2}
          }

   \institute{Instituto de Física y Astronomía,
              Universidad de Valparaíso, Gran Bretaña 1111, Valparaíso, Chile\\
              \email{graeme.candlish@uv.cl}
         \and
             Departamento de Física,
             Universidad Técnica Federico Santa María, Valparaíso, Chile\\
             \email{yara.jaffe@usm.cl}
             }

   \date{Received XXXX; accepted XXXX}
 
  \abstract{Given the mysterious nature of dark matter and dark energy, and the persistent tensions in cosmological data, it is worthwhile exploring more exotic physics in the dark sector, such as a momentum coupling between dark matter and dark energy, specifically in the form of a quintessence field. In this study, using collisionless N-body numerical simulations with a modified version of the RAMSES code, we follow up previous work to investigate the consequences of this model on dark matter halos and their substructures. We consider both the sign of the coupling and the imprints on structure formation and halo properties at a statistical level. We find that there is a clear enhancement (reduction) of substructure if the sign of the coupling is negative (positive) and that the dynamical state of the dark matter halos, particularly host halos, is undervirialised (overvirialised) at redshift zero when compared to uncoupled models or a reference $\Lambda$CDM simulation. Furthermore, positive coupling leads to less concentrated, less cuspy halos, whereas negative coupling leads to the opposite.}


   \maketitle

\section{Introduction}
The standard model of cosmology, referred to as $\Lambda$CDM \citep{Planck2018}, includes two dark components: cold dark matter (CDM) and dark energy in the form of a cosmological constant ($\Lambda$). Despite immense investigatory efforts, it remains unclear what the physical nature of these components may be. In the case of dark matter, many candidate particles have been proposed over the years (see \citealp{DM_review_2021} for a recent review), as well as the possibility that the phenomenology of dark matter is in fact explained by a modification of gravity in the weak-field limit (for a recent review of this approach, see \citealp{Banik_2022}). In the case of dark energy, the standard model assumes this to be represented by a cosmological constant; however, there are some very recent indications that dark energy may be dynamical and not in fact a cosmological constant \citep{DES_2025, DESI_2025}. Among the multitude of dynamical dark energy models, perhaps the most straightforward is that of quintessence \citep{Wetterich_1995, Caldwell_1998}, where a scalar field drives the late-time accelerated expansion.

An interesting extension of these models introduces a coupling between the two dark sector components \citep{Amendola2000, Wang2016}. There is also a multitude of possible forms that this coupling could take, with an appealing possibility being a momentum coupling between the dark matter component and the quintessence scalar field \citep{Pourtsidou_2013}. Several studies have investigated momentum-coupled models, sometimes referred to as dark scattering models \citep{Simpson_2010}, through the lens of numerical simulations \citep{Baldi&simpson_2015,Baldi&simpson_2017}. In our previous work reported in \citealp{Candlish2023}, a specific class of quintessence scalar field dark energy models with a momentum coupling to dark matter was investigated for the case of a positive coupling. In that study it was found that the coupling enhanced structure formation at larger scales, as measured by the power spectrum, but significantly suppressed structure on small scales, and reduced the slope of the inner density profiles of large mass halos. Furthermore, the velocities of  the substructure and the particle content of the halos was found to be significantly enhanced by the coupling. The evolution of the linear structure in these models has been studied in \citealp{Pourtsidou_2016,Chamings2020,SpurioMancini2021} with a view to resolving the   $\sigma_8$ tension \citep{Planck_2013_sz}.

In this work we extend our previous study in \citealp{Candlish2023} by considering higher resolution simulations, allowing us to analyse a much larger population of dark matter halos. In addition, we study the effect of a negative coupling between the dark sector components, something that was absent in our previous study. The particular focus of this study is to explore statistically the  effects of the coupling on halos and their substructure, motivated specifically by an interesting effect that was found in our earlier work whereby the velocity distributions of the most massive halos exhibited a bimodal behaviour, suggestive of a dynamically disturbed state, as is expected in cluster mergers. Given that the comparison simulations without the presence of the coupling showed that the same structure had fully virialised by redshift zero, our result suggested that the dynamical states of clusters may be significantly different in coupled models. However, the strength of this conclusion was severely limited by the low number statistics available due to the low resolution of our previous simulations. For this work we use higher resolution simulations to undertake a statistical study of this phenomenon.

\section{Theory}

We refer to our previous work \citealp{Candlish2023} and the original theoretical papers \citep{Pourtsidou_2013, Skordis_2015} for the full details of the theoretical background of the model we consider in this work. For a self-contained work, we here summarise the theoretical background to our study.

The energy-momentum tensor of the scalar field for Type 3 models is written as
\begin{equation}
    T^{(\phi)}_{\mu \nu} = F_Y \phi_\mu \phi_\nu - F g_{\mu \nu} - Z F_Z u_\mu u_\nu,
\end{equation}
where
\begin{equation}
\begin{split}
Y &= \frac{1}{2}\phi_\mu \phi^\mu\\
Z &= u^\mu \phi_\mu,
\end{split}
\end{equation}
and $F = F(Y,Z,\phi)$ is some function. $F_Y$ and $F_Z$ denote derivatives of this function with respect to $Y$ and $Z$, respectively. The four-velocity of the dark matter perfect fluid is given by $u^{\mu}$ and $\phi_\mu \equiv \partial_\mu \phi$. The equation of motion for the scalar field is given by
\begin{equation}
    \nabla_\mu (F_Y \phi^\mu + F_Z u^\mu) - F_\phi = 0,
\label{eom_phi}    
\end{equation}
while the dark matter fluid satisfies the usual equation for the conservation of density. The momentum transfer equation is given by
\begin{equation}
(\rho - ZF_Z)u^{\beta}\nabla_{\beta}u_{\mu} = \nabla_{\beta}(F_Zu^{\beta})\tilde{\phi}_{\mu} + F_ZD_{\mu} Z,
\label{momentum_eq}
\end{equation}
where $D_{\mu} = q^{\nu}_{\mu} \nabla_{\nu}$ is the spatial derivative operator given in terms of the projection operator $q^\nu _\mu \equiv u_{\mu}u^{\nu} + \delta_{\mu}^{\nu}$, and $\tilde{\phi}_{\mu} = q^{\nu}_{\mu} \nabla_{\nu} \phi = D_{\mu} \phi = \partial_{\mu} \phi + u^{\nu}u_{\mu} \partial_{\nu} \phi$ is the spatial projection of the derivative of the scalar field. We now specialise to the case of coupled quintessence by choosing $F = Y + V(\phi) + \gamma(Z)$. We work in the Newtonian gauge to facilitate passing to the non-relativistic limit. The line element in this gauge is given by
\begin{equation}
    ds^2 = a^2(\tau)[-(1+2\Psi)d\tau^2 + (1+2\Phi)\delta_{ij}dx^i dx^j],
\label{ds_perturbed}    
\end{equation}
where $\Phi$ and $\Psi$ are spatial scalars and $\delta_{ij}$ is the three-dimensional Kronecker delta. The perturbed fluid four-velocities (to linear order) in this gauge are
\begin{eqnarray*}
    u_0 &=& -a(1 + \Psi), \\
    u_i &=& a v_i.
\end{eqnarray*}
The evolution of the cold dark matter fluid at the background level and of the CDM fluid perturbations at linear level are given by the standard equations. However, the background evolution of the scalar field depends on the coupling and is given by equation~(\ref{eom_phi}) as
\begin{equation}
    \ddot{\phi} - \gamma_{ZZ}\ddot{\phi} + 2\mathcal{H}\dot{\phi} + \gamma_{ZZ}\mathcal{H}\dot{\phi} - 3a\mathcal{H}\gamma_Z + a^2V_\phi = 0.
\label{scalar_field_background}
\end{equation}
From equation (\ref{eom_phi}) at first order in the perturbations we obtain
\begin{equation}
\begin{split}
V_{\phi \phi} \varphi a^2 &+ 3\gamma_Z \Psi a \mathcal{H} + 3\gamma_Z a \dot{\Phi} - \gamma_Z a \nabla^2 \theta - \gamma_{ZZZ} \frac{\ddot{\bar{\phi}}}{a} \dot{\varphi}\\ 
&+ \gamma_{ZZZ} \frac{\dot{\bar{\phi}}}{a} \dot{\varphi} \mathcal{H} + 2\gamma_{ZZ}\Psi \ddot{\bar{\phi}} - 2\gamma_{ZZ}\Psi \dot{\bar{\phi}} \mathcal{H} + \gamma_{ZZ}\dot{\Psi}\dot{\bar{\phi}} \\
&- \gamma_{ZZ}\ddot{\varphi} - 2\gamma_{ZZ} \dot{\varphi} \mathcal{H} - 2\Psi \ddot{\bar{\phi}} - 4\Psi\dot{\bar{\phi}}\mathcal{H} - 3\dot{\Phi}\dot{\bar{\phi}} - \dot{\Psi}\dot{\bar{\phi}} + \ddot{\varphi} \\
&- \nabla^2 \varphi + 2\dot{\varphi} \mathcal{H} = 0.
\label{scalar_field_perturbation}
\end{split}
\end{equation}
Passing to Fourier space and taking the Newtonian limit (non-relativistic velocities are implicit in the gauge choice as the DM fluid velocity perturbation is assumed to satisfy $|v| \ll 1$) of $k\gg\mathcal{H}$, this equation simplifies drastically to 
\begin{equation}
\label{eq:varphi}
\varphi = a\gamma_z\theta.
\end{equation}
With this we write the momentum transfer equation (\ref{momentum_eq}) as
\begin{equation}
\begin{split}
\dot{\theta} + \mathcal{H}\theta + \Psi  = & \frac{1}{a\bar{\rho} - \gamma_Z \dot{\bar{\phi}}} \left[2\gamma_Z \dot{\bar{\phi}} \theta \mathcal{H} + 3a\gamma_Z^2 \theta \mathcal{H} - \gamma_Z \Psi \dot{\bar{\phi}} \right. \\
&\left. + \gamma_Z \ddot{\bar{\phi}} \theta + a \gamma_Z^2 \mathcal{H} \theta + a \gamma_Z \gamma_{ZZ} \dot{\bar{Z}} \theta + a \gamma_Z^2 \dot{\theta} \right] \\
- & \frac{1}{a^2 \bar{\rho} - a \gamma_Z \dot{\bar{\phi}}} \left[\gamma_{ZZ}\dot{\bar{\phi}}^2 \theta \mathcal{H} - \mathcal{H} \gamma_Z \gamma_{ZZ} \dot{\bar{\phi}} \theta \right. \\
&\left. + \gamma_{ZZ} \dot{\bar{\phi}} \ddot{\bar{\phi}} \theta + \gamma_Z \gamma_{ZZ} \dot{\bar{\phi}} \theta \right],
\end{split}
\label{momentum_transfer_eq_2}
\end{equation}
where $\dot{\bar{Z}} = 1/a(\ddot{\phi} + \mathcal{H}\dot{\phi})$. We now must choose a form for the coupling, and so we follow \cite{Pourtsidou_2013} and choose
\begin{equation}
    \gamma(Z) = \gamma_0 Z^2
    \label{eq:gammaZ},
\end{equation}
where $\gamma_0$ is a constant whose value is assumed to be in the range $0 \leq \gamma_0 < 1/2$. The equation (\ref{momentum_transfer_eq_2}) becomes
\begin{equation}
(1+h_1) \dot{v}_i  + (1+h_2) \mathcal{H}v_i + (1+h_3)\nabla_i \Psi = 0
\label{eq:modified_euler_eq}
\end{equation}
where the coefficients $h_1$, $h_2$, and $h_3$ are
\begin{equation}
\begin{split}
h_1 &= \frac{4\gamma_0^2 \dot{\phi}^2}{a^2\rho - 2\gamma_0\dot{\phi^2}}, \\
h_2 &= \frac{ (8\gamma_0^2 - 2\gamma_0)\dot{\phi}^2 + (8\gamma_0^2 - 4\gamma_0)\dot{\phi}\ddot{\phi}  \frac{1}{\mathcal{H}}}{ a^2\rho - 2\gamma_0\dot{\phi}^2},\\
h_3 &= \frac{ 2\gamma_0\dot{\phi}^2}{ a^2\rho - 2\gamma_0\dot{\phi}^2}.
\end{split}
\label{h_vals}
\end{equation}
In the $h_2$ term, we can replace $\ddot{\phi}$ using the evolution equation of the background field, given by equation (\ref{scalar_field_background}), with 
\begin{equation}
\ddot{\phi}(1-2\gamma_0) + 2\mathcal{H}\dot{\phi}(1-2\gamma_0) + a^2 V_{\phi} = 0.
\label{scalar_field_background2},
\end{equation}
where we have used equation (\ref{eq:gammaZ}). Thus, $h_2$ can be written as \begin{equation}
h_2 = \frac{ 4\gamma_0 (\frac{3}{2} - 2\gamma_0)\dot{\phi}^2 + 4\gamma_0\dot{\phi} a^2 V_{\phi}/\mathcal{H}}{a^2\rho - 2\gamma_0\dot{\phi}^2}.
\end{equation}
Now that we have the modified Euler equation, we must now implement this in our cosmological N-body code. We use the well-known RAMSES code \citep{Teyssier_2002}, which is a grid-based adaptive mesh refinement (AMR) code, using a particle-mesh (PM) scheme to evolve the dark matter particle distribution. Our implementation begins with revisiting the   supercomoving coordinates \citep{Martel_1998} used in RAMSES, which are defined as
\begin{equation}
\begin{split}
\vec{v} &= H_0 L \frac{1}{a}\tilde{\vec{u}}, \\
\vec{x} &= \frac{1}{a}\frac{\tilde{\vec{x}}}{L}, \\
dt &= a^2 \frac{d\tilde{t}}{H_0}, \\
\Psi &= \frac{L^2H_0^2}{a^2} \tilde{\Phi},
\end{split}
\label{coord_supercomoviles}
\end{equation}
where $L$ is the length of the simulation box. The coordinates denoted with a tilde are the supercomoving coordinates. To simplify the notation, we apply the transformation and then remove the tildes. Thus, using Eq. (\ref{coord_supercomoviles}) in Eq. (\ref{eq:modified_euler_eq}) we get
\begin{equation}
    \frac{d\vec{u}}{dt} = -\frac{h_2 - h_1}{1 + h_1}a^2 \frac{H}{H_0} \vec{u} - \frac{1+h_3}{1+h_1} \vec{\nabla} \Phi.
    \label{euler_ec_standardform}
\end{equation}
We now have the modified Euler Eq. (\ref{euler_ec_standardform}) in a form in which it may be discretised and solved numerically. To connect with the implementation of the modified Euler equation in the code, we write the finite difference update of the velocity, as implemented in RAMSES using the Leapfrog scheme, in the following manner:
\begin{equation}
    \frac{v^{n+1/2}_p - v^{n}_p}{(1/2)\Delta t^n} = F.
\end{equation}
Here $F$ is the force acting on the particle. We can now modify the velocity update as required to implement equation (\ref{euler_ec_standardform}) in the following way:
\begin{equation}
    v^{n+1/2}_p = v^n_p - \frac{h_2 - h_1}{1+h_1}a^2\frac{H}{H_0}v^n_p \Delta t^n/2 + \frac{1+h_3}{1+h_1}F \Delta t^n/2.
\label{vel_timestep_hvalues}    
\end{equation}
We define two new coefficients, $\epsilon_1$ and $\epsilon_2$, to simplify the expression:
\begin{equation}
\begin{split}
\epsilon_1 &= 1 - \frac{h_2-h_1}{1+h_1}a^2\frac{H}{H_0}\Delta t^n/2 ,\\
\epsilon_2 &= \frac{1+h_3}{1+h_1}.
\end{split}
\end{equation}
Thus, finally Eq. (\ref{vel_timestep_hvalues}) becomes
\begin{equation}
        v^{n+1}_p = \epsilon_1 v^n_p + \epsilon_2 F \Delta t^n/2.
\label{eq_ramses}        
\end{equation}
This is the equation we have implemented in RAMSES. The standard dynamics is recovered by setting $\epsilon_1 = \epsilon_2 = 1$, which is equivalent to having all the $h_i$ equal to zero.

\section{Simulations}

   \begin{figure}
   \centering
   \includegraphics[width=0.5\textwidth]{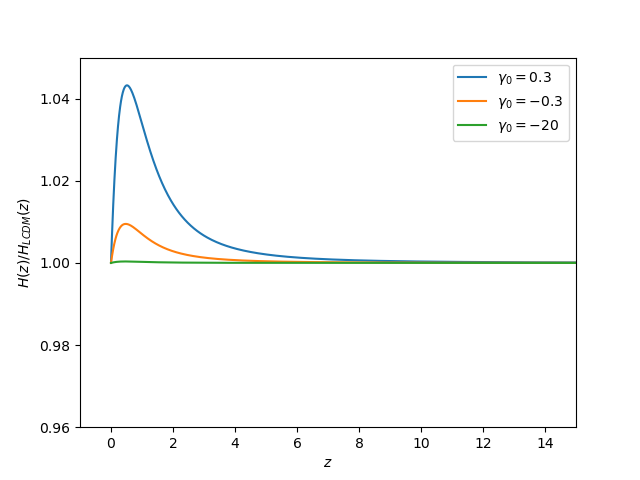}
      \caption{Hubble parameter normalised by that of $\Lambda$CDM for the three values of the coupling constant.
              }
         \label{Fig:hubble}
   \end{figure}

   \begin{figure}
   \centering
   \includegraphics[width=0.5\textwidth]{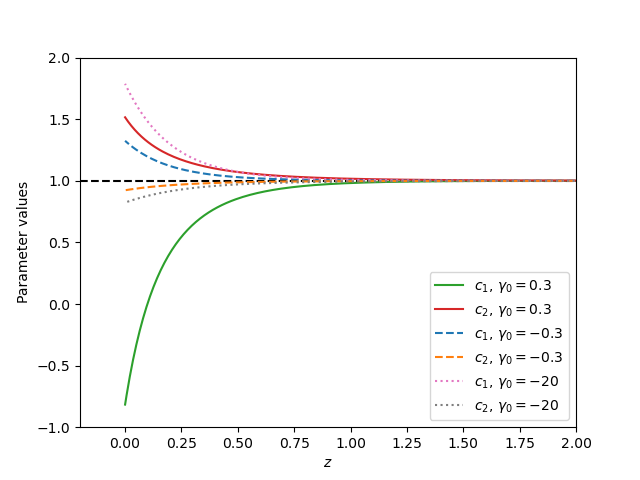}
      \caption{Modified Euler equation coefficients for the three values of the coupling constant.
              }
         \label{Fig:euler}
   \end{figure}

We focus on the same scalar field model, referred to as model C, as used in our previous work. This model has a potential of the form of \citealp{Albrecht_2000}, given by
\begin{equation}
    V(\phi) = ((\phi-\beta)^{\alpha}+\Gamma)e^{-\lambda \phi},
\end{equation}
where the parameters are chosen to be $\beta = 3.8$, $\Gamma = 20.0$, $\alpha = 17.0$, and $\lambda = 0.934$. The last value is fixed by demanding that the closure condition of the energy densities is satisfied (assuming spatial flatness). The initial values of the scalar field and its derivative are $\phi = 100$, $\dot{\phi} = 10$. In \citealp{Albrecht_2000} the form of this potential is justified as providing a model where all parameters are roughly of order unity in Planck units, that inherits useful properties from the well-known tracker solutions of exponential models, and that deviates from a pure exponential in the correct manner to provide the appropriate late-time acceleration. Furthermore, potentials of this form are expected to emerge naturally in the low energy limit of string theory-inspired models. For our purposes, this model was chosen primarily based on the phenomenological constraint of closely matching to the Hubble constant evolution of $\Lambda$CDM in order to reproduce the observational successes of that model.

We generated the initial conditions for our simulations using the MonofonIC code \citep{Monofonic} (sometimes referred to as MUSIC2). This code incorporates the CLASS code for the generation of the initial transfer function. We  modified the CLASS code used in order to incorporate the coupling between the dark matter and the scalar field in order to generate consistent initial conditions, although the effect of the coupling is minimal at high redshift. The cosmological parameters that we used are those of \citealp{Planck2018}, with the obvious exception that we did not use $\Omega_{\Lambda}$ but a quintessence field. The parameters are $\Omega_b = 0.0494$, $\Omega_{\text{CDM}} = 0.264979$, $\Omega_{\phi} = 0.68412$, $h = 0.67321$, $\sigma_8 = 0.8102$, and $n_s = 0.9661$.

We note that the presence of the coupling leads to an overall scaling of the background potential for the scalar field:
\begin{equation}
\ddot{\phi} + 2\mathcal{H}\dot{\phi} + \frac{1}{1-2\gamma_0}a^2V_{\phi} = 0.
\end{equation}
Therefore, the background evolution is modified by changing the value of the coupling constant. In order to separate the effects of the coupling on structure formation due to the modified dark matter dynamics as compared to simply a modified Hubble parameter, we ran simulations (as in our previous study) where the background evolution is modified by the coupling, but the Euler equation is fixed to the standard form (i.e. the coupling is switched off for the particle dynamics).

The background evolution of the Hubble parameter for the coupled models is given in Fig~\ref{Fig:hubble}, as compared to $\Lambda$CDM. It is apparent that there is a more pronounced and longer deviation from $\Lambda$CDM seen in the case of positive coupling, while for small negative coupling the deviation is much reduced, and for large negative coupling it is very close to $\Lambda$CDM. It should be noted that the positive coupling model is affected by a strong coupling problem at $\gamma_0 = 1/2$, and so it is to be expected that values of $\gamma_0$ closer to that critical value  exhibit more pronounced effects. In any case, the deviations from the standard model do not exceed $\sim$$4\%$. In the case of a negative coupling there is no theoretical restriction in the allowed value of the coupling constant.

We also show the evolution of the parameters $c_1$ and $c_2$ in the modified Euler equation in Fig.~\ref{Fig:euler}. We can see that a positive coupling leads to a significant enhancement of the gravitational acceleration term $c_2$, which we would naively expect to lead to an increase in small-scale structure in this model. As we have shown before, this is in fact not the case. The reason is that the cosmological friction term $c_1$ for positive coupling leads to an effective push (which we call the cosmological push, as opposed to cosmological friction) where the particles are further accelerated beyond what is expected from the gravitational effects alone. For negative coupling, we have the opposite behaviour: the cosmological friction is now enhanced, while the gravitational acceleration is reduced. We note that the change in the effective gravitational force (the $c_2$ term) is not as pronounced as the change in the effective cosmological friction (the $c_1$ term) when comparing the difference between the two negatively coupled models. The $c_2$ coefficient is approximately $12\%$ smaller for $\gamma_0 = -20$ as compared to $\gamma_0 = -0.3$, whereas the $c_1$ coefficient is approximately $35\%$ larger for $\gamma_0 = -20$ as compared to $\gamma_0 = -0.3$. Again we might naively expect to see a reduction in structure due to the increased relevance of the cosmological friction term, but in fact we see, at small scales, an enhancement of structure.

All of our simulations use $512^3$ particles in a $128^3$ Mpc h$^{-1}$ volume, leading to a particle mass of order $1.3 \times 10^9$ M$_{\odot}$ h$^{-1}$. We set the minimum particle number for a dark matter halo at $20$, and thus we do not resolve any halos less massive than approximately $2.6 \times 10^{10}$ M$_{\odot}$ h$^{-1}$. The minimum grid resolution is $250$ kpc h$^{-1}$ and the maximum refined resolution is $7.8$ kpc h$^{-1}$. The full list of simulations that we use is given in Table~\ref{table:sims}.

\begin{table}
\caption{\label{table:sims}Simulation names and associated coupling values.}
\centering
\begin{tabular}{ l c c c}
 \hline\hline 
Simulation Name & $\gamma_0$ & Coupled? & $G$ value\\
 \hline
 g03 & $0.3$ & Yes & Effective \\
 gm03 & $-0.3$ & Yes & Effective \\
 gm20 & $-20.0$ & Yes & Effective \\
 g03\_u & $0.3$ & No & Effective \\
 gm03\_u & $-0.3$ & No & Effective \\
 gm20\_u & $-20.0$ & No & Effective\\
 g03\_G & $0.3$ & Yes & Newtonian \\
 gm20\_G & $-20.0$ & Yes & Newtonian \\
 $\Lambda$CDM & NA & NA & Newtonian \\
 \hline
\end{tabular}
\tablefoot{Models with the coupling deactivated for the particle dynamics are indicated with a `u' in their name for "uncoupled". The last two coupled simulations are identical to the g03 and gm20 models, but we use the Newtonian gravitational constant in the halo analysis.}
\end{table}

\section{Results}

\subsection{Density projections}

   \begin{figure*}
   \centering
   \includegraphics[width=\textwidth]{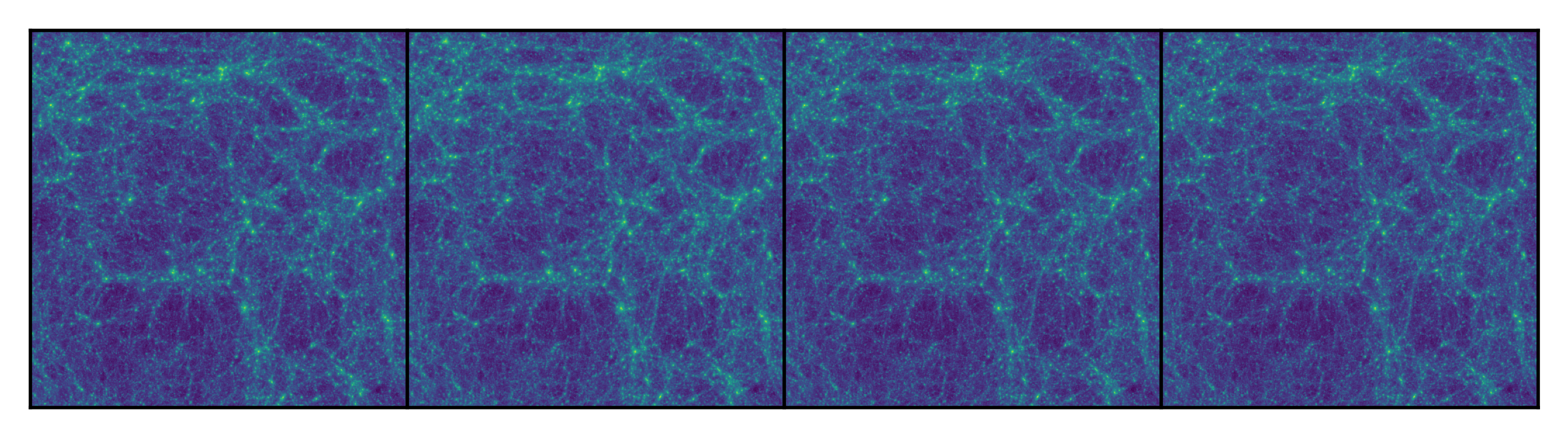}
      \caption{Density projection plots of all simulations considered in this work. The leftmost panel is our $\Lambda$CDM run, and the other three panels (from left to right) are coupled models with: $\gamma_0 = 0.3$, $\gamma_0 = -0.3$, $\gamma_0 = -20$.
              }
         \label{fig:density}
   \end{figure*}

As a first simple check of our results, we confirmed that the overall density field of the structure formed in our simulations is visually identical. We did this using simple density projection plots where the particles are binned into a two-dimensional pixel mesh of $512 \times 512$ pixels. Each particle contributes in the same way to the total density within each pixel (i.e. the density calculation is unweighted). These results are shown in Fig.~\ref{fig:density}. It is clear, at the qualitative level of a visual inspection of these density plots, that all models generate structure in mostly the same locations within the computational box, although perhaps there is some indication of differing degrees of density contrast, as indicated by the colour associated with the value of the projected particle density within that pixel. This is to be expected as the coupling affects the degree of structure formation that takes place, as we  see from the examination of the power spectra of our simulations in the next section.

\subsection{Power spectrum}

   \begin{figure}
   \centering
   \includegraphics[width=0.5\textwidth]{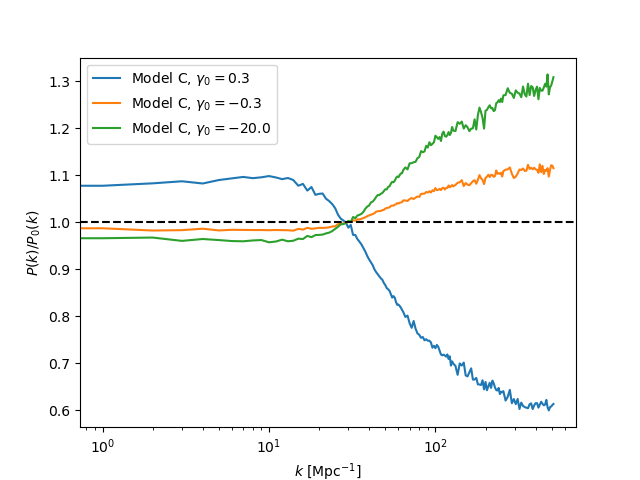}
      \caption{Power spectra normalised by the corresponding uncoupled model.
              }
         \label{Fig:PowSpec}
   \end{figure}

Then we analysed the power spectra of our coupled models, normalised with respect to the associated uncoupled model. In other words, we divided the power spectrum of model p0.3 by that of p0.3\_u, the power spectrum of model m0.3 by that of m0.3\_u, and finally the power spectrum of m20.0 by that of m20.0\_u. This was to extract the effect of the coupling from the modified background evolution in each model. In all cases the power spectra were calculated using the POWMES code \citep{Colombi_2009}.

In Fig.~\ref{Fig:PowSpec} we plot the normalised power spectra. We can see that the positive coupling model behaves, as expected, in the same manner as we found in our earlier analysis described in \citealp{Candlish2023}: on large linear scales we see a mostly scale-independent increase in power, while at small non-linear scales there is a pronounced (scale-dependent) reduction in power  compared to the uncoupled model. In our previous study the loss of small-scale power was understood to be due to the significantly modified dynamics of the dark matter whereby high velocities (such as those found within a major overdensity) are further enhanced by what we   refer  to as the cosmological push. This push removes dark matter from the overdense regions, thus reducing structure. An interesting aspect of this phenomenon is that the modification of the velocity vector due to this effect is proportional to the Hubble parameter and the velocity vector itself, rather than being proportional to the gravitational potential gradient. Thus, the dynamics induced by the cosmological push are less directly connected to the density distribution, leading to late-time dynamical states of the substructure that exhibit less virialisation than is expected in a non-coupled model. We   explore this much more fully in the following sections.

In the case of a negatively coupled model we see opposite behaviour: the large-scale linear power spectrum is slightly suppressed  compared to the coupled model, while the small-scale non-linear power spectrum shows a strong enhancement of power, particularly for the $\gamma_0 = -20$ model, although it is a less pronounced effect than the reduction in power shown in the positively coupled model. Inverting the argument given above for positive coupling, we expect that the enhanced cosmological friction acts to inhibit structure formation in a scale-independent manner, thus leading to large-scale suppression of structure formation. At small scales, within overdensities, the particle velocities are enhanced, leading to a more pronounced friction effect, which (despite the reduced gravitational acceleration) is sufficient to reduce velocities such that more bound virialised structures are formed. We note that the positive coupling model shows more pronounced deviations from the uncoupled model due to the coupling $\gamma_0 = 0.3$ being close to the strong coupling value of $\gamma_0 = 1/2$ where singular behaviour arises in the equations of motion.

\subsection{Halo mass function and velocity dispersions}

\begin{figure}
\centering
\includegraphics[width=0.5\textwidth]{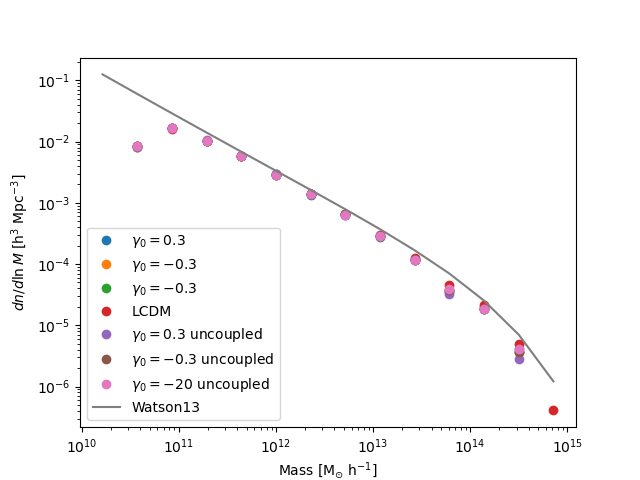}
\caption{Halo mass functions for all simulations considered in this study. The comparison line given is from \citealp{Watson_2013}, as defined in the Colossus Python package.}
\label{fig:hmf}
\end{figure}

   \begin{figure}
   \centering
   \includegraphics[width=0.5\textwidth]{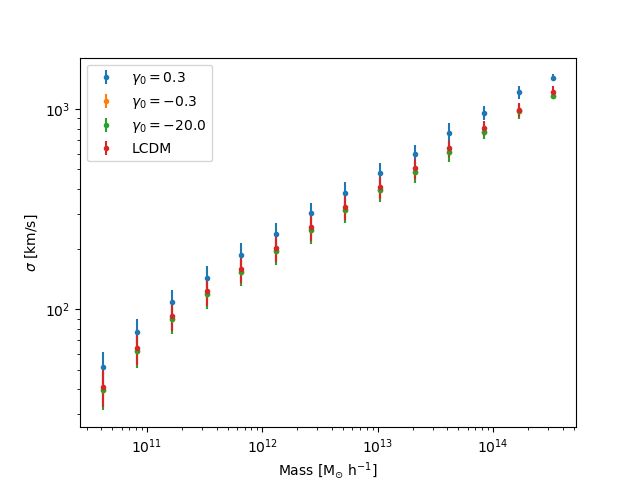}
      \caption{Velocity dispersion as a function of mass for all halos in all the coupled models considered in this study and the reference $\Lambda$CDM model.}
         \label{fig:vel_disp}
   \end{figure}

The halo catalogues from our simulation runs were generated by using the Amiga Halo Finder (AHF) \citep{Knollmann_2009}. This uses an overdensity criterion to identify which particles belong to which halos. In addition, the escape velocity of each particle is calculated and multiplied by a factor of $1.5$ to determine whether the particle is actually gravitationally bound to the halo. We note that in our analysis in our prior study \citep{Candlish2023}, we modified the effective Newtonian gravitational constant used in the calculation of the escape velocity according to the value of the $c_2$ coefficient. We did this again for this study, where the effective values of the gravitational constant (at $z=0$) are given in Table \ref{table:grav}. We   also considered   how the analysis is affected if we assume the Newtonian value of the gravitational constant, supposing that in an observational context we would not be privy to the effective gravitational constant experienced by the dark matter. For the halo mass function, however, the difference between using these effective gravitational constants and using the standard Newtonian value are negligible. We used the conventional value of $\Delta_{200}$ (i.e. $200$ times the background density) to define the maximum radius of our halos.

\begin{table}
\caption{\label{table:grav}Effective gravitational constants for the coupled models}
\centering
\begin{tabular}{ l c c}
 \hline\hline 
$\gamma_0$ & $G'$ [Mpc km$^2$ M$_{\odot}^{-1}$ s$^{-2}$] \\
 \hline
 $0.3$ & $6.52 \times 10^{-9}$ \\
 $-0.3$ & $3.97 \times 10^{-9}$ \\
 $-20.0$ & $3.54 \times 10^{-9}$ \\
 \hline
\end{tabular}
\tablefoot{$G' = c_2G$ where $c_2$ is the coefficient of the gravitational acceleration in the modified Euler equation (evaluated at $z=0$) and $G$ is the standard Newtonian value.}
\end{table}

The halo mass functions for our simulations are shown in Fig.~\ref{fig:hmf}, with a comparison to the \citealp{Watson_2013} fitting line extracted from a suite of very large particle number N-body simulations. We can see that the halo counts in the intermediate mass bins compare reasonably well with the reference line, except at the low and high mass extremes, as expected, where insufficient mass resolution and insufficient box size, respectively, limit our results. It is worth noting that the presence or otherwise of the coupling, and the sign of the coupling, have a negligible impact on the halo mass function.

To begin our analysis of the velocities of the particles within the dark matter halos, we plot in Fig.~\ref{fig:vel_disp} the velocity dispersions, as measured by AHF, for halos of all masses found in the simulations. Within each mass bin we determine the standard deviation of the measured velocity dispersions, which are indicated by the vertical error bars. We immediately see that the positive coupling model exhibits a clear increase in the velocity dispersions across all mass bins, as compared to all other models. The negatively coupled models show slightly suppressed velocity dispersions as compared to $\Lambda$CDM but within the errors they are identical. The uncoupled models (not plotted) are also identical to $\Lambda$CDM within the errors. We note that the velocity dispersions are also not significantly sensitive to the use of the effective gravitational constant as opposed to the Newtonian value.

\subsection{Bimodality in the velocity distributions}

\subsubsection{Bimodality parameter}

   \begin{figure}
   \centering
   \includegraphics[width=0.5\textwidth]{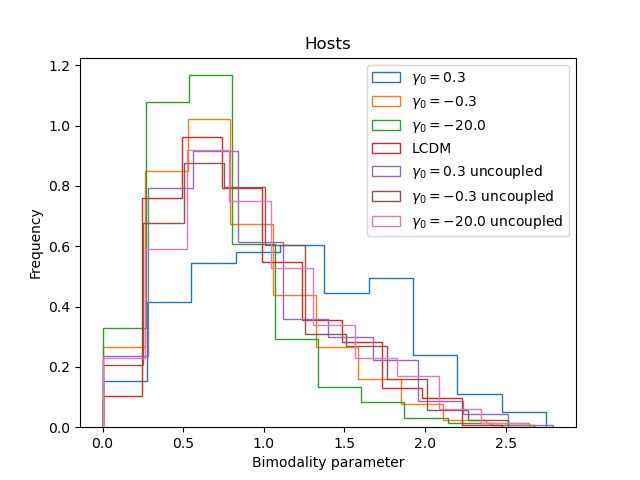}
      \caption{Histogram of bimodality parameter values calculated from the bi-Gaussian distribution fitted to the velocity magnitude histograms of all particle velocities in the $500$ most massive host halos of each simulation.}
         \label{Fig:Bimodal_hosts}
   \end{figure}

   \begin{figure}
   \centering
   \includegraphics[width=0.5\textwidth]{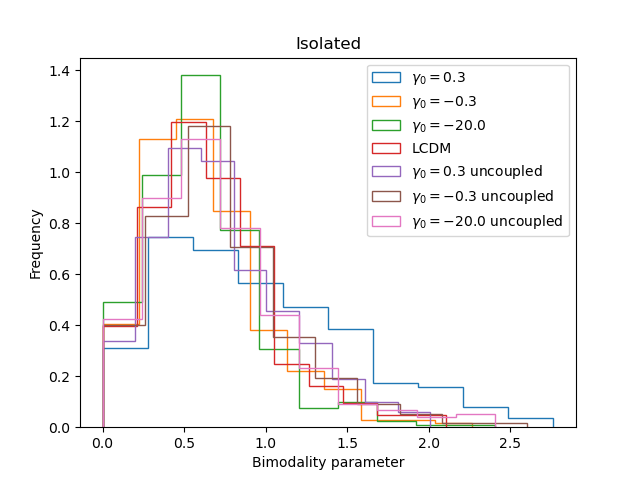}
      \caption{As in Fig.~\ref{Fig:Bimodal_hosts}, but for isolated halos.}
         \label{Fig:Bimodal_isolated}
   \end{figure}

   \begin{figure}
   \centering
   \includegraphics[width=0.5\textwidth]{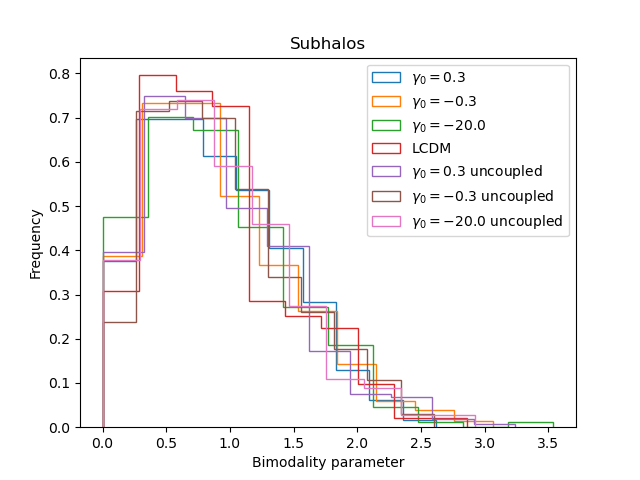}
      \caption{As in Fig.~\ref{Fig:Bimodal_hosts}, but for subhalos.}
         \label{Fig:Bimodal_subhalos}
   \end{figure}

\begin{figure}
\centering
\includegraphics[width=0.5\textwidth]{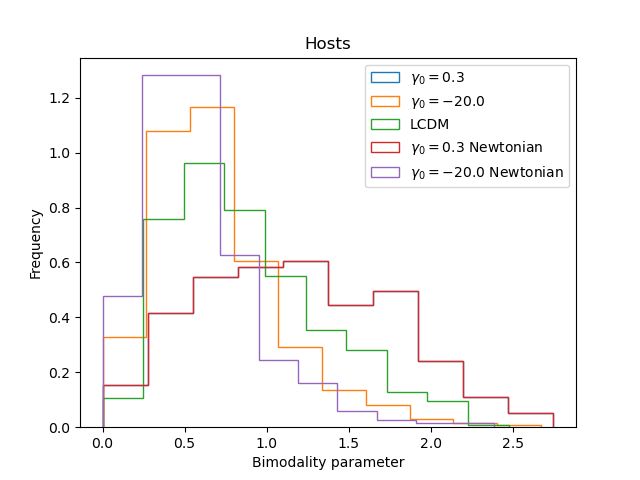}
\caption{As in Fig. \ref{Fig:Bimodal_hosts}, but using the Newtonian gravitational constant in the halo finding process.}
\label{Fig:Bimodal_hosts_Newton}
\end{figure}

\begin{figure}
\centering
\includegraphics[width=0.5\textwidth]{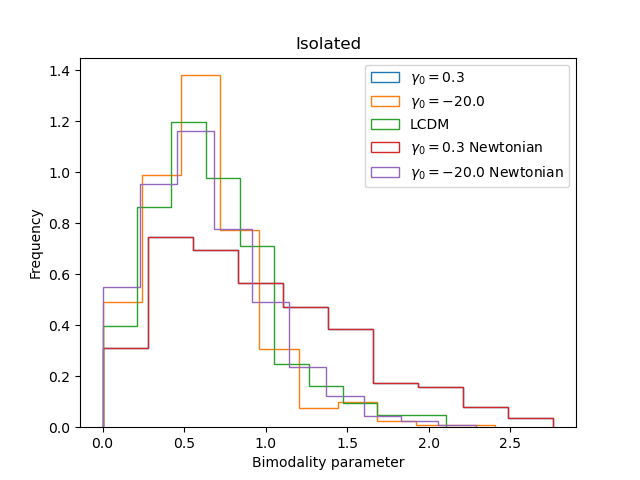}
\caption{As in Fig. \ref{Fig:Bimodal_isolated}, but using the Newtonian gravitational constant in the halo finding process.}
\label{Fig:Bimodal_isolated_Newton}
\end{figure}

\begin{figure}
\centering
\includegraphics[width=0.5\textwidth]{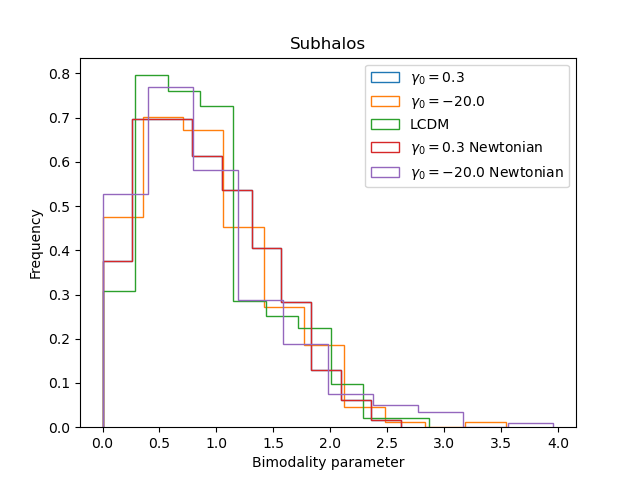}
\caption{As in Fig. \ref{Fig:Bimodal_subhalos}, but using the Newtonian gravitational constant in the halo finding process.}
\label{Fig:Bimodal_subhalos_Newton}
\end{figure}

As discussed earlier, in previous work we found a strikingly bimodal velocity distribution for the most massive halo in the positively coupled model, indicating a halo that has yet to reach a fully virialised state. This was, however, an effect seen in only a single halo. For this work we     studied this phenomenon in much more detail, and from a statistical perspective given the much higher number of halos available in our present simulations.

We analysed the degree of bimodality in the following manner. Firstly, we separate the halos into three categories: host, isolated, or subhalo. A host halo is simply any halo that contains at least one other halo, as determined by AHF. This category thus includes large clusters as well as small groups where a single subhalo is within a host. A subhalo is a halo that is contained within another halo. Finally, an isolated halo is a halo that is neither a host nor a subhalo.

After this categorisation, for each of the $500$ most massive halos within each category in our simulations, we generate the histogram of 3D velocity magnitudes (normalised by each halo's velocity dispersion) for all particles within the halo, using $60$ evenly spaced bins. We then fit to that histogram the following bimodal Gaussian distribution:
\begin{equation}
    f(v) = A_1\exp\left(-\frac{(v-\mu_1)^2}{2\sigma_1}\right) + A_2\exp\left(-\frac{(v-\mu_2)^2}{2\sigma_2}\right).
\end{equation}
Here $v$ denotes the $3D$ velocity magnitude, the $\mu_i$ values are the means of each component, the $A_i$ values are the amplitudes, and the $\sigma_i$ values are the standard deviations. We compare the values of $A_1$ and $A_2$ to determine which of the two Gaussian components has the larger amplitude. The parameters of the larger Gaussian is referred to as $A_L$, $\mu_L$, and $\sigma_L$, while the parameters of the smaller Gaussian will be referred to as $A_S$, $\mu_S$, and $\sigma_S$.

We then define the following bimodality parameter:
\begin{equation}
    B = \frac{A_S}{A_L}\frac{|\mu_S - \mu_L|}{\sigma_L}.
\end{equation}
The motivation for this parameter is to quantify the degree of bimodality of the velocity magnitude histograms. The factor of the amplitude ratios will reduce the value of this parameter towards zero if the smaller Gaussian is significantly smaller than the larger, as in this case the smaller Gaussian would be essentially irrelevant to the distribution. The absolute value of the difference between the mean values is a straightforward measure of the bimodality of the double Gaussian distribution, which we measure in units of the standard deviation of the larger Gaussian. We find that this parameter is an effective measure of the bimodality in almost all cases, except for the smallest subhalos, where the relatively low particle number leads to noisy histograms, which causes some of the larger amplitude Gaussian components to be assigned very small standard deviations, thus leading to very high bimodality parameter values. For the subhalos we therefore limit the histograms to values of $B < 5$.

In Fig.~\ref{Fig:Bimodal_hosts} we plot the histograms of the bimodality parameters for the $500$ most massive host halos in each simulation. Here we see a clear distinction between the positive coupling model (blue) and all other models, where the host halos exhibit a clear bimodality, and relatively infrequent unimodal distributions. This confirms the preliminary results found in our previous study \citep{Candlish2023}. The bimodality found in a single massive host halo in that study has been found to be prevalent throughout the host halo population of the positively coupled model in the present work.

The uncoupled models and $\Lambda$CDM all show broadly similar behaviour, as does the small negative coupling model, with a preference towards unimodal distributions (small parameter values) and a tail of bimodality. Interestingly, however, in the model with a large negative coupling (green) the preference towards unimodal velocity distributions is considerably stronger, with bimodality far less frequent when compared to the uncoupled models (or $\Lambda$CDM). This suggests that, statistically, halos in this model are more virialised than either the uncoupled models or $\Lambda$CDM.

Now we turn our attention to the isolated halos in Fig. \ref{Fig:Bimodal_isolated}. We can see in this case that for most of the models, particularly $\Lambda$CDM and the small negatively coupled model, bimodality is relatively scarce in these halos; instead, we see a statistical preference towards unimodal distributions. This is to be expected given that these are isolated halos that  either  have not undergone any merger or have not undergone a recent merger such that any subhalo has fully dissolved within the host. Interestingly, this contrasts with the positive coupling model where there is still an increased probability of finding a bimodal distribution, even though these halos are isolated. This would suggest that the material stripped during some past merger is still undergoing virialisation. In addition, the large negative coupling model shows a slightly reduced probability for bimodality and an increased likelihood of unimodal distributions.

Finally, the subhalos are given in Fig. \ref{Fig:Bimodal_subhalos}. Given that these halos are undergoing tidal interactions within a host halo, and are likely to be far from virialised, we  thus expect some degree of bimodality, and we do indeed see a larger tail of higher bimodality parameters in this case. This behaviour is, however, clearly independent of the coupling.

In the previous analysis of the bimodality, all halos are calculated using the effective gravitational constants as listed in Table \ref{table:grav}. Clearly we know what these values must be from our simulations. However, it is worthwhile to explore how these results are modified if we instead assume a position of ignorance and work with the standard Newtonian value of $G$, as would be the case in a hypothetical observational analysis.

In Figs. \ref{Fig:Bimodal_hosts_Newton}, \ref{Fig:Bimodal_isolated_Newton}, and \ref{Fig:Bimodal_subhalos_Newton} we compare the coupled models with $\gamma_0 = 0.3$ and $\gamma_0 = -20$ using the modified gravitational constant and those models using the Newtonian value. We also include the $\Lambda$CDM results for reference. By using the Newtonian gravitational constant in the halo finding procedure for these models we are modifying which particles are considered bound to each halo. In the positive coupling case the effective gravitational constant is higher than the Newtonian value, thus using the lower Newtonian value in the halo finding process could lead to the unbinding of high velocity particles which are otherwise included. In the case of a negative coupling, the effective gravitational constant is lower, thus using the higher Newtonian value may lead to unbound high velocity particles as being considered bound.

In Fig. \ref{Fig:Bimodal_hosts_Newton} we compare only the host halos, which, as we have seen, exhibit the greatest degree of bimodality. We see that the histogram for the positively coupled model using the Newtonian value is identical to that seen by using the effective gravitational constant. Thus, any unbound particles have no significant effect on the degree of bimodality. In the negatively coupled case we see that the unimodality of the velocity distribution is even more pronounced when using the Newtonian gravitational constant (purple line) as opposed to the effective gravitational constant (orange line). For the isolated halos in Fig. \ref{Fig:Bimodal_isolated_Newton} we see again that the bimodality of the positively coupled model is unaffected by the choice of gravitational constant. However, The negatively coupled model again shows some difference:  the use of the Newtonian gravitational constant leads to a slight reduction in the unimodality of these halos. In either case, the isolated halos in the negatively coupled model exhibit more unimodality than bimodality. Finally, in Fig. \ref{Fig:Bimodal_subhalos_Newton} we see that the subhalo distributions are essentially unaffected by the gravitational constant used.

\subsubsection{Probability of bimodality}

   \begin{figure}
   \centering
   \includegraphics[width=0.5\textwidth]{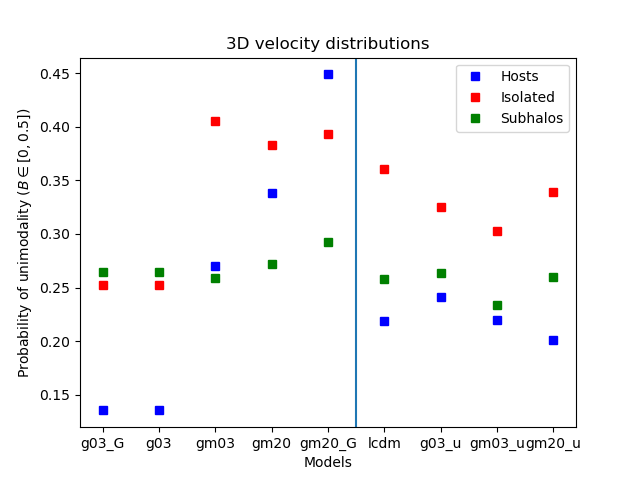}
      \caption{Probabilities of unimodality in 3D velocity distributions for all models.}
         \label{Fig:probs_uni_3d}
   \end{figure}

   \begin{figure}
   \centering
   \includegraphics[width=0.5\textwidth]{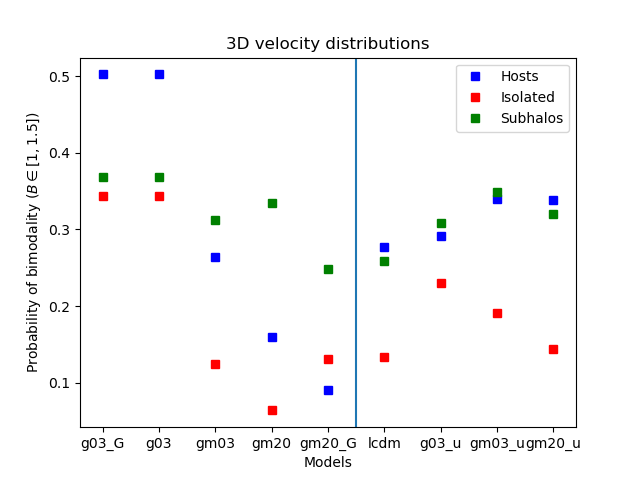}
      \caption{Probabilities of bimodality in 3D velocity distributions for all models.}
         \label{Fig:probs_bi_3d}
   \end{figure}

   \begin{figure}
   \centering
   \includegraphics[width=0.5\textwidth]{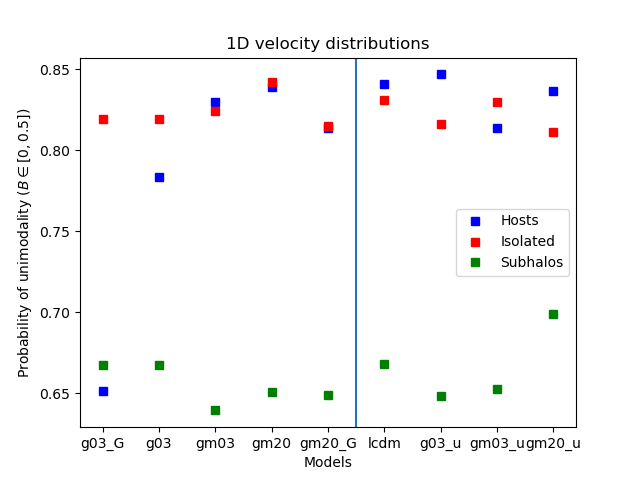}
      \caption{Probabilities of unimodality in 1D velocity distributions for all models.}
         \label{Fig:probs_uni_1d}
   \end{figure}

\begin{figure}
\centering
\includegraphics[width=0.5\textwidth]{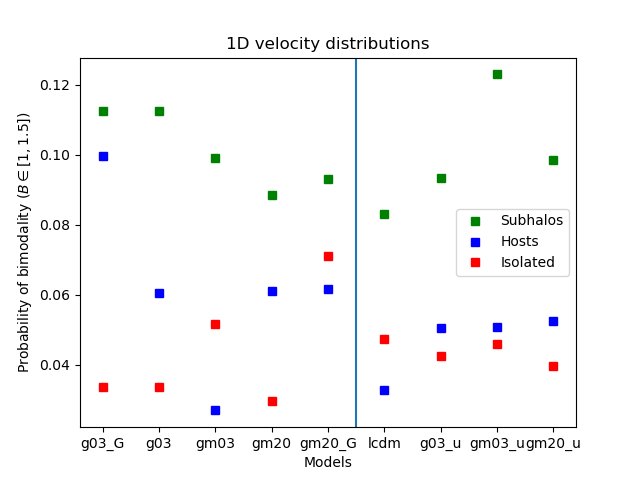}
\caption{Probabilities of bimodality in 1D velocity distributions for all models.}
\label{Fig:probs_bi_1d}
\end{figure}

We further quantify the degree of unimodality or bimodality by determining the probabilities of observing certain ranges of bimodality parameters for our models. We limit the bimodality distributions for all halo types to $B=3$. We then calculate the areas under our histograms for certain ranges of the bimodality parameter. The first range we consider is values of $B$ from $0$ to $0.5$, which we consider to be indicative of a unimodal distribution. The other range we consider is values of $B$ from $1$ to $2$, which we consider to be indicative of a bimodal distribution. Furthermore, to connect our work more closely with observationally accessible quantities, we consider 1D velocity distributions by repeating our bimodal Gaussian fitting procedure with only the $v_x$ component of the particle velocity vectors.

The results of the 3D analysis are shown in Figs.~\ref{Fig:probs_uni_3d}, \ref{Fig:probs_bi_3d}, while the probabilities obtained from the 1D bimodality parameter histograms are given in Figs.~\ref{Fig:probs_uni_1d} and \ref{Fig:probs_bi_1d}. The coupled models are separated from the uncoupled models and $\Lambda$CDM by a vertical blue line to aid in reading these plots. 

For the case of unimodality in the 3D velocity distributions (Fig.~\ref{Fig:probs_uni_3d}) we see  for the host halos that there is a clear separation as we move from the positively coupled model (with either the effective or Newtonian gravitational constant) to the small negative coupling, large negative coupling, and then large negative coupling with the Newtonian gravitational constant. The probability of unimodal host distributions is below $15\%$ for the positively coupled model, whereas it is around $35\%$ for the large negatively coupled model (or $45\%$ if using a Newtonian analysis). For the isolated halos, there is a higher probability for unimodality in negatively coupled models, around $40\%$,  compared to $25\%$ for the positively coupled model. The subhalo unimodality probability is roughly $25\%-30\%$ for all coupled models. In the uncoupled cases and $\Lambda$CDM the host halos have a unimodal probability of $\sim 25\%$, similarly for subhalos, while the isolated halos have a somewhat higher probability of unimodality,  $\sim 35\%$.

In Fig.~\ref{Fig:probs_bi_3d} we consider the probability of bimodality. First we look at host halos. For the positively coupled model this probability is high, at $\sim 50\%$, whereas for the large negative coupling model the probability is low, around $15\%$ (or below $10\%$ in a Newtonian analysis). Isolated halos in the positively coupled model have $\sim 35\%$ probability of bimodality, while this is far lower for isolated halos in all the negatively coupled models, being  $\sim 10\%$. For the uncoupled and $\Lambda$CDM models the hosts and subhalos have $\sim 30\%$ probability of bimodality, whereas the isolated halos have a probability of $\sim 20\%$ (with rather large variation across uncoupled models).

The trends clearly seen in the 3D velocity distributions become much less clear when using 1D velocity distributions, as  expected. In the case of unimodality (Fig.~\ref{Fig:probs_uni_1d}) for the host halos, the positively coupled model, using the Newtonian analysis, has a probability of $65\%$, whereas it is $\sim 78\%$ using the effective gravitational constant. For the negatively coupled models, the uncoupled models and $\Lambda$CDM the probability of unimodality is around $80\%-85\%$. The isolated halos all show a similar probability, regardless of coupling. The subhalos for all models show a probability of a unimodal distribution of around $65\%-70\%$. Thus, for the 1D velocity distributions only the positively coupled model using a Newtonian analysis is clearly discriminated.

For the case of bimodality, again only the positively coupled model in a Newtonian analysis shows a clear difference, with a bimodality probability in the host halos of $\sim 10\%$, whereas it lies at $6\%$ or below for all other models. Isolated halos and subhalos appear to offer no means for model discrimination in the case of 1D velocities.

\subsection{Halo properties}

\subsubsection{Mass, radius, and concentration}

   \begin{figure}
   \centering
   \includegraphics[width=0.5\textwidth]{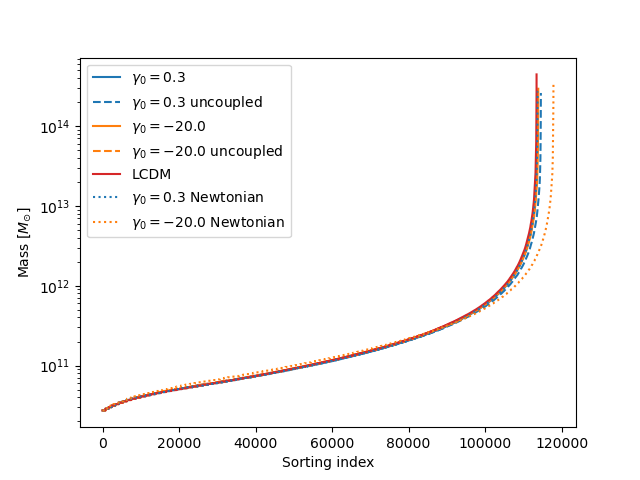}
      \caption{Sorted halo masses for all halos in each model.}
         \label{Fig:sorted_masses}
   \end{figure}

   \begin{figure}
   \centering
   \includegraphics[width=0.5\textwidth]{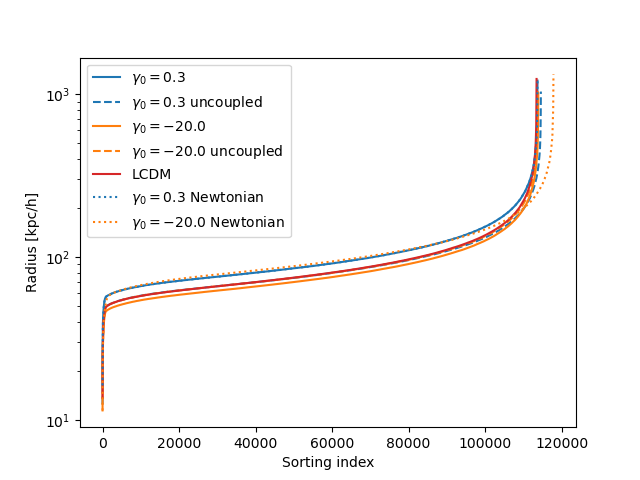}
      \caption{Sorted halo virial radii for all halos in each model.}
         \label{Fig:sorted_radii}
   \end{figure}

   \begin{figure}
   \centering
   \includegraphics[width=0.5\textwidth]{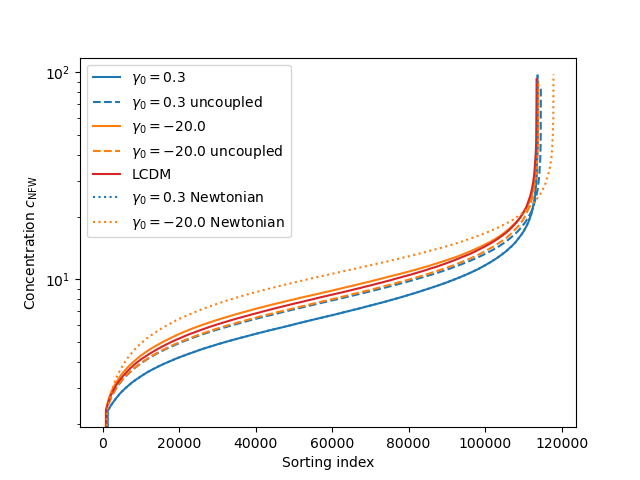}
      \caption{Sorted halo concentrations for all halos in each model.}
         \label{Fig:sorted_concentrations}
   \end{figure}

The halo masses, radii, and concentrations are compared across the entire halo catalogues in Figs. \ref{Fig:sorted_masses}, \ref{Fig:sorted_radii}, and \ref{Fig:sorted_concentrations}. In these plots we   sort the halo catalogues of each model according to each property and then plot these sorted values together, with the sorting index referring simply to the halo number according to the chosen ordering. Due to differences in the number of halos in each model, the curves do not necessarily overlap at the extreme right of the plots, where we see a shift in the horizontal position of the curves. If the distributions of properties are broadly similar, however, we   see a significant overlap of the curves for all index values except the highest and, in particular, we do not see any noticeable shift in the vertical positions of the curves.

For the masses of the halos we cannot distinguish the curves, except for the horizontal shift at high sorting index due to differing halo numbers, as discussed earlier. This is consistent with the similarity in the halo mass functions of all models. This tells us that halo mass is not an effective means of discriminating between models, and that modification of the gravitational constant does not break the degeneracy in this property. 

For the radii, however, we see a vertical shift for the positively coupled model, indicating that the halos in this model are more extended than in the uncoupled models. The halo radii of the negatively coupled model, when using the appropriate effective gravitational constant, are slightly below those of the uncoupled models, indicating less radially extended halos. This difference is eliminated when working with the Newtonian gravitational coupling, indicating that gravitationally unbound particles at large radii are included in these halos. Given the overlap of the $\gamma_0 = -20$ curve (using the Newtonian value of $G$) and the $\gamma_0 = 0.3$ curve, we conclude that halo radius is not an effective means of discriminating between models if the gravitational constant is assumed Newtonian.

Finally, the halo concentrations are given in Fig. \ref{Fig:sorted_concentrations}. In this case the uncoupled models are coincident, but all coupled models are discriminated, with a clear vertical separation in their curves. The negative coupling leads to more concentrated halos, while the positive coupling leads to less concentrated halos. The separation is even more stark when working with the Newtonian gravitational constant, leading to significantly higher values for the concentration parameter in the negatively coupled model.

\subsubsection{Derivative of the density profile}

   \begin{figure}
   \centering
   \includegraphics[width=0.5\textwidth]{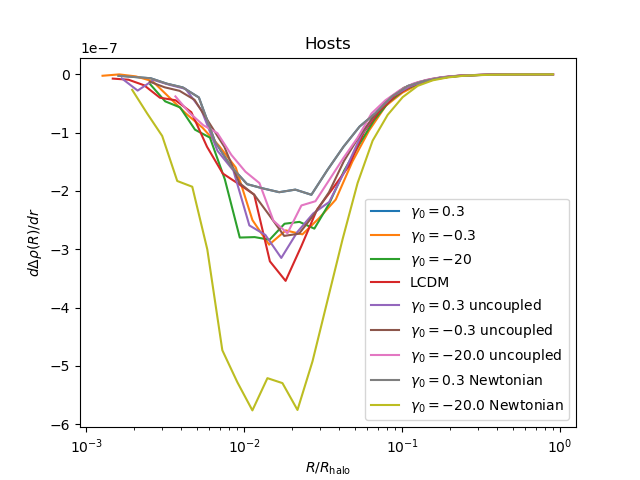}
      \caption{Averaged density contrast derivative for the first $500$ halos in each model.}
         \label{Fig:density_deriv}
   \end{figure}

In Fig.~\ref{Fig:density_deriv} we show the derivative of the averaged density contrast profile for the first $500$ halos in each simulation, with the radii normalised by the virial radius of each halo, to ensure these values range from $0$ to $1$. We now explain the details of how we build this plot. Rather than work with the density itself, we consider the density contrast (i.e. the density in units of the background density value). Furthermore, we are interested in the slope of the density profile, and thus we calculate the derivative of the profile in a simple way: we calculate the difference in the density contrast values between each radial bin divided by the differences in the normalised radii between each bin. These derivatives are then averaged over all $500$ halos. We note that the radii are plotted on a logarithmic scale in Fig. \ref{Fig:density_deriv}.

We find that there is a difference in the inner slope, at around $1\%-10\%$ of the halo virial radius, depending on the presence or otherwise of coupling, and the sign and magnitude of that coupling. Firstly, we note that the lowest values are found for the positively coupled model (regardless of the gravitational constant used) indicating, on average, less steep inner density profiles for this model. The results for the other models are broadly similar, showing steeper inner profiles in the negatively coupled models, mostly consistent with those of the $\Lambda$CDM reference simulation. It is also very clear that the inner profile is drastically steepened if the Newtonian gravitational constant is used in the halo analysis, suggesting that the halo finding process in this case includes a substantial population of additional bound particles in the inner regions of these halos.

\subsubsection{Velocity dispersion profile}

   \begin{figure}
   \centering
   \includegraphics[width=0.5\textwidth]{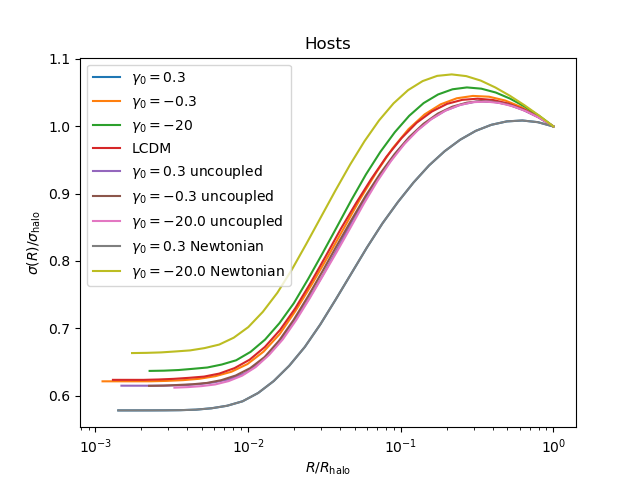}
      \caption{Averaged velocity dispersion profiles for the first $500$ halos in each model.}
         \label{Fig:vel_div_profiles}
   \end{figure}

We now stack the velocity dispersion profiles for the first $500$ halos in each model, in the same manner as we did in the previous section for the density profile derivatives. The velocity dispersions are normalised by $\sigma(R_{\text{halo}})$, leading to a profile value of unity at $R_{\text{halo}}$. The results of this analysis are given in Fig.~\ref{Fig:vel_div_profiles}, where we see that the Newtonian velocity dispersions are either substantially larger (smaller) at all radii for negative (positive) coupling. A less pronounced radius-independent enhancement is seen for the negatively coupled model using the effective gravitational constant;  the other models are mostly indistinguishable. It is worth noting, however, that the peak of the normalised velocity dispersion profiles shifts towards smaller radii, relative to $R_{\text{halo}}$, as the coupling changes from positive to negative, with this being most pronounced again for the Newtonian analysis.

\subsubsection{Virial ratios}

\begin{figure}
   \centering
   \includegraphics[width=0.5\textwidth]{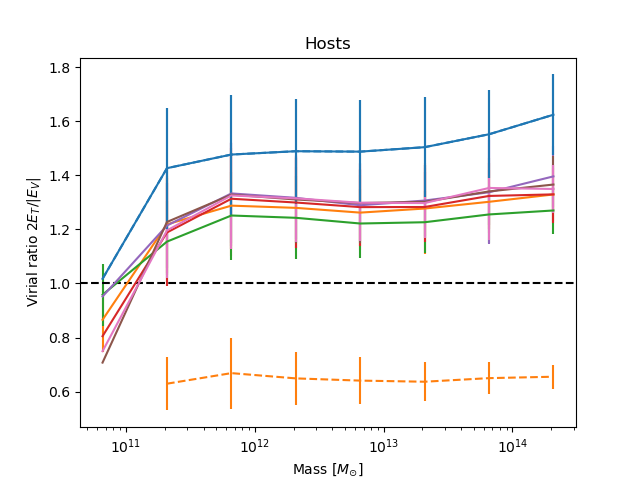}
      \caption{Virial ratios for the host halos, binned in mass. The error bars show the standard deviation of the virial ratios within each bin.}
         \label{Fig:host_vir}
   \end{figure}

\begin{figure}
   \centering
   \includegraphics[width=0.5\textwidth]{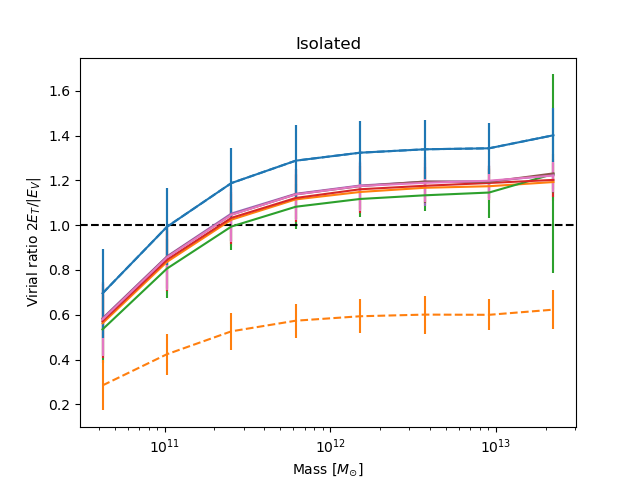}
      \caption{Virial ratios for the isolated halos, binned in mass. The error bars show the standard deviation of the virial ratios within each bin.}
         \label{Fig:isolated_vir}
   \end{figure}

\begin{figure}
   \centering
   \includegraphics[width=0.5\textwidth]{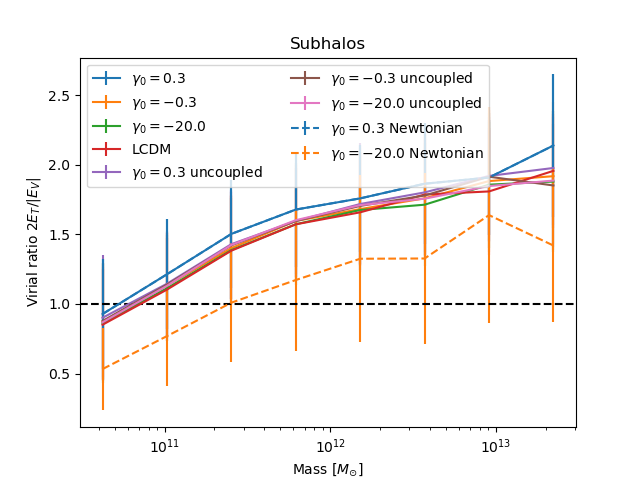}
      \caption{Virial ratios for the subhalos, binned in mass. The error bars show the standard deviation of the virial ratios within each bin.}
         \label{Fig:subhalo_vir}
   \end{figure}

In Figs.~\ref{Fig:host_vir}, \ref{Fig:isolated_vir} and \ref{Fig:subhalo_vir} we plot the halo virial ratios (determined using all halo particles) for hosts, isolated halos, and subhalos, respectively. These results are also binned in mass.

We find that the virial ratios for the hosts, Fig. \ref{Fig:host_vir}, for all the uncoupled models, and for our reference $\Lambda$CDM model are broadly similar and indicate somewhat overvirialised halos for all masses except the least massive. A similar picture emerges for the isolated halos and the subhalos. The isolated halos are closer to being virialised, as expected, while the subhalos, especially those that are more massive, are substantially overvirialised, most likely due to tidal interactions within their hosts.

The positively coupled model, indicated with the blue line in all three plots, shows a clear positive vertical offset in the virial ratios of their halos, across all masses, especially for the hosts. The external tidal interactions are expected to be dominant for the subhalos, leading to a virialisation behaviour, which is mostly consistent with that seen for the other models.

When analysed using the effective gravitational constant arising from the coupling, the host halos in the negatively coupled model show a slight reduction in their virial ratios, with a behaviour consistent with the uncoupled models for the isolated halos and the subhalos. If we work with the Newtonian gravitational constant, however, almost all halos in the negatively coupled model are significantly undervirialised. This is especially the case for the isolated halos. The subhalos show a trend that is consistent with the other models, but with lower values across all masses of the virial ratio.

This analysis is consistent with the bimodalities seen earlier: the positive coupling leads to dynamically disturbed clusters and groups, and to isolated halos, as compared to $\Lambda$CDM, whereas the negative coupling (at least for $\gamma_0 = -20$) leads to these halos being more dynamically cold than   expected in the standard model. This effect is even more pronounced if the gravitational constant assumed in the analysis is the Newtonian value.

\section{Conclusions}
In this study we have extended previous work to analyse the effect of a coupling between dark matter and dark energy in the form of a quintessence scalar field, by focusing on the halo properties in a statistical manner. We have specifically concentrated on the dynamical state of the halos, particularly the hosts, confirming earlier results of a bimodality in the velocity distribution of the halo particles. This may be a general signal of positively coupled models, with an associated super-virial dynamical state. Furthermore, we have extended our work to negatively coupled models, showing a signature tendency towards unimodal velocity distributions and sub-virial halos. In addition, kinematical halo properties such as the steepness of the inner density profile and the halo concentration may help discriminate between negatively and positively coupled models. The halo mass-concentration relation has previously been shown to be a useful probe of the interaction strength in coupled models \citep{Zhao2023}. We now list our specific conclusions.

   \begin{itemize}
      \item There is a very significant reduction ($40\%$) in structure at smaller scales for positive coupling that is sufficiently close to the value $\gamma_0 = 1/2$, as seen in our previous work. For negative coupling there is a significant enhancement (up to $30\%$ for the larger coupling) in structure at smaller scales. On linear scales this behaviour is inverted. This is consistent with other studies of momentum-coupled (or dark scattering) models \citep{Simpson_2010,Baldi&simpson_2015,Baldi&simpson_2017}.
      \item The dynamical state of the dark matter halos, particularly host halos, is more super-virial than seen in the uncoupled models and is more likely to exhibit bimodality in their particle velocities in the case of positive coupling. For (large) negative coupling the opposite is the case: host halos are somewhat less super-virial than seen in the uncoupled models and much less likely to exhibit bimodality in their particle velocities. 
      \item The results regarding the halo dynamical states for negative coupling is even more pronounced if the analysis assumes that the dark matter dynamics are Newtonian. Then the host and isolated halos are significantly sub-virial, with very clear unimodality in the velocity distributions of most of the host halos.
      \item In the case of positive coupling the halo concentrations are reduced  compared to uncoupled models and $\Lambda$CDM. For negative coupling the halo concentrations are increased  compared to uncoupled models and $\Lambda$CDM, especially if the analysis assumes Newtonian gravitational accelerations.
      \item Consistent with the effect on the halo concentrations, we find that the slope of the inner density profiles is reduced (compared to the uncoupled models and $\Lambda$CDM) in the positive coupling case, whereas it is increased in the case of negative coupling, especially so in a Newtonian analysis.
   \end{itemize}

The central conclusion of our work is that using averaged information from the full halo population, both  kinematical information about the halos (halo concentration and slope of the inner density profile) and information about their dynamical states (bimodality and virialisation), allows us to discriminate the sign and magnitude of the coupling in these models. The discriminatory effectiveness of these properties is even stronger if the analysis is undertaken assuming Newtonian gravitational accelerations, as would be the case in an observational analysis, where a standard value of $G$ would be taken as a minimal prior.

There are several possible extensions to our study that would be worthwhile to pursue. First, it would be beneficial to run hydrodynamical simulations in order to connect these results with the baryonic component, and thus open the door to a direct comparison with observations. In this vein, one possible line of investigation would be to consider very high resolution zoom simulations of individual galaxies or galaxy groups embedded in a cosmological context in the presence of these couplings. In addition, as always, it would be helpful to have higher resolution simulations to be able to explore these effects in lower mass halos. From a theoretical point of view, it would be interesting to further explore the parameter space of these models by considering larger negative coupling values, as well as other potentials for the scalar field, giving rise to alternative background evolutions. In this context, it would be of interest to explore models where the coupling has relevance at higher redshift, perhaps tending to the standard behaviour at late times. Generalisations of these models are also possible, such as considering other forms for the coupling term, or even mixing a momentum coupling with a density coupling.

The signatures of these models discussed in this study may eventually present an opportunity to test these ideas with observational data, allowing us to further constrain the vast space of possible cosmological models.

\begin{acknowledgements}
      GNC wishes to thank Alkistis Pourtsidou for very useful discussions.
\end{acknowledgements}

\bibliographystyle{aa}
\bibliography{mibiblio}

\begin{thebibliography}{27}
\expandafter\ifx\csname natexlab\endcsname\relax\def\natexlab#1{#1}\fi

\bibitem[{Ade {et~al.}(2014)Ade, Aghanim, Armitage-Caplan, Arnaud, Ashdown, Atrio-Barandela, Aumont, Baccigalupi, Banday, \& et~al.}]{Planck_2013_sz}
Ade, P. A.~R., Aghanim, N., Armitage-Caplan, C., {et~al.} 2014, Astronomy \& Astrophysics, 571, A20

\bibitem[{Albrecht \& Skordis(2000)}]{Albrecht_2000}
Albrecht, A. \& Skordis, C. 2000, Physical Review Letters, 84, 2076–2079

\bibitem[{Amendola(2000)}]{Amendola2000}
Amendola, L. 2000, Physical Review D, 62

\bibitem[{{Arbey} \& {Mahmoudi}(2021)}]{DM_review_2021}
{Arbey}, A. \& {Mahmoudi}, F. 2021, Progress in Particle and Nuclear Physics, 119, 103865

\bibitem[{{Baldi} \& {Simpson}(2015)}]{Baldi&simpson_2015}
{Baldi}, M. \& {Simpson}, F. 2015, \mnras, 449, 2239

\bibitem[{{Baldi} \& {Simpson}(2017)}]{Baldi&simpson_2017}
{Baldi}, M. \& {Simpson}, F. 2017, \mnras, 465, 653

\bibitem[{{Banik} \& {Zhao}(2022)}]{Banik_2022}
{Banik}, I. \& {Zhao}, H. 2022, Symmetry, 14, 1331

\bibitem[{Caldwell {et~al.}(1998)Caldwell, Dave, \& Steinhardt}]{Caldwell_1998}
Caldwell, R.~R., Dave, R., \& Steinhardt, P.~J. 1998, Physical Review Letters, 80, 1582–1585

\bibitem[{{Chamings} {et~al.}(2020){Chamings}, {Avgoustidis}, {Copeland}, {Green}, \& {Pourtsidou}}]{Chamings2020}
{Chamings}, F.~N., {Avgoustidis}, A., {Copeland}, E.~J., {Green}, A.~M., \& {Pourtsidou}, A. 2020, \prd, 101, 043531

\bibitem[{{Colombi} {et~al.}(2009){Colombi}, {Jaffe}, {Novikov}, \& {Pichon}}]{Colombi_2009}
{Colombi}, S., {Jaffe}, A., {Novikov}, D., \& {Pichon}, C. 2009, \mnras, 393, 511

\bibitem[{{DES Collaboration} {et~al.}(2025){DES Collaboration}, {Abbott}, {Acevedo}, {Adamow}, {Aguena}, {Alarcon}, {Allam}, {Alves}, {Andrade-Oliveira}, {Annis}, {Armstrong}, {Avila}, {Bacon}, {Bechtol}, {Blazek}, {Bocquet}, {Brooks}, {Brout}, {Burke}, {Camacho}, {Camilleri}, {Campailla}, {Carnero Rosell}, {Carr}, {Carretero}, {Castander}, {Cawthon}, {Chan}, {Chang}, {Chen}, {Conselice}, {Costanzi}, {Crocce}, {da Costa}, {Pereira}, {Davis}, {De Vicente}, {Deiosso}, {Desai}, {Diehl}, {Dodelson}, {Doux}, {Drlica-Wagner}, {Elvin-Poole}, {Everett}, {Ferrero}, {Fert{\'e}}, {Flaugher}, {Frieman}, {Galbany}, {Garc{\'\i}a-Bellido}, {Gatti}, {Gaztanaga}, {Giannini}, {Gruen}, {Gruendl}, {Gutierrez}, {Hartley}, {Herner}, {Hinton}, {Hollowood}, {Honscheid}, {Huterer}, {James}, {Jeffrey}, {Jeltema}, {Kessler}, {Lahav}, {Lee}, {Lee}, {Lidman}, {Lin}, {Lin}, {Marshall}, {Mena-Fern{\'a}ndez}, {Miquel}, {Muir}, {M{\"o}ller}, {Nichol}, {Palmese}, {Paterno}, {Percival}, {Pieres}, {Plazas Malag{\'o}n}, {Popovic}, {Porredon},
  {Prat}, {Qu}, {Raveri}, {Rodriguez-Monroy}, {Romer}, {Rykoff}, {Sako}, {Samuroff}, {Sanchez}, {Sanchez Cid}, {Scolnic}, {Sevilla-Noarbe}, {Shah}, {Sheldon}, {Smith}, {Suchyta}, {Sullivan}, {Swanson}, {S{\'a}nchez}, {Tarle}, {Taylor}, {Thomas}, {To}, {Toribio San Cipriano}, {Toy}, {Troxel}, {Tucker}, {Vikram}, {Vincenzi}, {Walker}, {Weaverdyck}, {Weller}, {Wiseman}, {Yamamoto}, \& {Yanny}}]{DES_2025}
{DES Collaboration}, {Abbott}, T.~M.~C., {Acevedo}, M., {et~al.} 2025, arXiv e-prints, arXiv:2503.06712 (to be submitted to PRD)

\bibitem[{{DESI Collaboration} {et~al.}(2025){DESI Collaboration}, {Abdul-Karim}, {Aguilar}, {Ahlen}, {Alam}, {Allen}, {Allende Prieto}, {Alves}, {Anand}, {Andrade}, {Armengaud}, {Aviles}, {Bailey}, {Baltay}, {Bansal}, {Bault}, {Behera}, {BenZvi}, {Bianchi}, {Blake}, {Brieden}, {Brodzeller}, {Brooks}, {Buckley-Geer}, {Burtin}, {Calderon}, {Canning}, {Carnero Rosell}, {Carrilho}, {Casas}, {Castander}, {Cereskaite}, {Charles}, {Chaussidon}, {Chaves-Montero}, {Chebat}, {Chen}, {Claybaugh}, {Cole}, {Cooper}, {Cuceu}, {Dawson}, {de la Macorra}, {de Mattia}, {Deiosso}, {Della Costa}, {Demina}, {Dey}, {Dey}, {Ding}, {Doel}, {Edelstein}, {Eisenstein}, {Elbers}, {Fagrelius}, {Fanning}, {Fern\'andez-Garc\'ia}, {Ferraro}, {Font-Ribera}, {Forero-Romero}, {Frenk}, {Garcia-Quintero}, {Garrison}, {Gaztañaga}, {Gil-Mar\'in}, {Gontcho}, {Gonzalez}, {Gonzalez-Morales}, {Gordon}, {Green}, {Gutierrez}, {Guy}, {Hadzhiyska}, {Hahn}, {He}, {Herbold}, {Herrera-Alcantar}, {Ho}, {Honscheid}, {Howlett}, {Huterer}, {Ishak}, {Juneau},
  {Kamble}, {Kara\c\{c\}ayl\{\i\}}, {Kehoe}, {Kent}, {Kim}, {Kirkby}, {Kisner}, {Koposov}, {Kremin}, {Krolewski}, {Lahav}, {Lamman}, {Landriau}, {Lang}, {Lasker}, {Le Goff}, {Le Guillou}, {Leauthaud}, {Levi}, {Li}, {Li}, {Lodha}, {Lokken}, {Lozano-Rodr\'iguez}, {Magneville}, {Manera}, {Martini}, {Matthewson}, {Meisner}, {Mena-Fern\'andez}, {Menegas}, {Mergulh\~ao}, {Miquel}, {Moustakas}, {Muñoz-Guti\'errez}, {Muñoz-Santos}, {Myers}, {Nadathur}, {Naidoo}, {Napolitano}, {Newman}, {Niz}, {Noriega}, {Paillas}, {Palanque-Delabrouille}, {Pan}, {Peacock}, {Pellejero Ibanez}, {Percival}, {P\'erez-Fern\'andez}, {P\'erez-R\`afols}, {Pieri}, {Poppett}, {Prada}, {Rabinowitz}, {Raichoor}, {Ram\'irez-P\'erez}, {Rashkovetskyi}, {Ravoux}, {Rich}, {Rocher}, {Rockosi}, {Rohlf}, {Rom\'an-Herrera}, {Ross}, {Rossi}, {Ruggeri}, {Ruhlmann-Kleider}, {Samushia}, {Sanchez}, {Sanders}, {Schlegel}, {Schubnell}, {Seo}, {Shafieloo}, {Sharples}, {Silber}, {Sinigaglia}, {Sprayberry}, {Tan}, {Tarl\'e}, {Taylor}, {Turner}, {Ureña-L\'opez},
  {Vaisakh}, {Valdes}, {Valogiannis}, {Vargas-Magaña}, {Verde}, {Walther}, {Weaver}, {Weinberg}, {White}, {Wolfson}, {Y\`eche}, {Yu}, {Zaborowski}, {Zarrouk}, {Zhai}, {Zhang}, {Zhao}, {Zhao}, {Zhou}, \& {Zou}}]{DESI_2025}
{DESI Collaboration}, {Abdul-Karim}, M., {Aguilar}, J., {et~al.} 2025, arXiv e-prints, arXiv:2503.14738 (part of DR2 publication series)

\bibitem[{{Hahn} {et~al.}(2020){Hahn}, {Michaux}, {Rampf}, {Uhlemann}, \& {Angulo}}]{Monofonic}
{Hahn}, O., {Michaux}, M., {Rampf}, C., {Uhlemann}, C., \& {Angulo}, R.~E. 2020, {MUSIC2-monofonIC: 3LPT initial condition generator}, Astrophysics Source Code Library, record ascl:2008.024

\bibitem[{Knollmann \& Knebe(2009)}]{Knollmann_2009}
Knollmann, S.~R. \& Knebe, A. 2009, The Astrophysical Journal Supplement Series, 182, 608–624

\bibitem[{Martel \& Shapiro(1998)}]{Martel_1998}
Martel, H. \& Shapiro, P.~R. 1998, Monthly Notices of the Royal Astronomical Society, 297, 467–485

\bibitem[{{Palma} \& {Candlish}(2023)}]{Candlish2023}
{Palma}, D. \& {Candlish}, G.~N. 2023, \mnras, 526, 1904

\bibitem[{{Planck Collaboration} {et~al.}(2018){Planck Collaboration}, {Aghanim}, {Akrami}, {Ashdown}, {Aumont}, {Baccigalupi}, {Ballardini}, {Banday}, {Barreiro}, {Bartolo}, \& et~al.}]{Planck2018}
{Planck Collaboration}, {Aghanim}, N., {Akrami}, Y., {et~al.} 2018, Astronomy \& Astrophysics, 641, A6

\bibitem[{Pourtsidou {et~al.}(2013)Pourtsidou, Skordis, \& Copeland}]{Pourtsidou_2013}
Pourtsidou, A., Skordis, C., \& Copeland, E.~J. 2013, Physical Review D, 88

\bibitem[{Pourtsidou \& Tram(2016)}]{Pourtsidou_2016}
Pourtsidou, A. \& Tram, T. 2016, Physical Review D, 94

\bibitem[{{Simpson}(2010)}]{Simpson_2010}
{Simpson}, F. 2010, \prd, 82, 083505

\bibitem[{Skordis {et~al.}(2015)Skordis, Pourtsidou, \& Copeland}]{Skordis_2015}
Skordis, C., Pourtsidou, A., \& Copeland, E. 2015, Physical Review D, 91

\bibitem[{{Spurio Mancini} \& {Pourtsidou}(2022)}]{SpurioMancini2021}
{Spurio Mancini}, A. \& {Pourtsidou}, A. 2022, \mnras, 512, L44

\bibitem[{{Teyssier}(2002)}]{Teyssier_2002}
{Teyssier}, R. 2002, \aap, 385, 337

\bibitem[{Wang {et~al.}(2016)Wang, Abdalla, Atrio-Barandela, \& Pavón}]{Wang2016}
Wang, B., Abdalla, E., Atrio-Barandela, F., \& Pavón, D. 2016, Reports on Progress in Physics, 79, 096901

\bibitem[{{Watson} {et~al.}(2013){Watson}, {Iliev}, {D'Aloisio}, {Knebe}, {Shapiro}, \& {Yepes}}]{Watson_2013}
{Watson}, W.~A., {Iliev}, I.~T., {D'Aloisio}, A., {et~al.} 2013, \mnras, 433, 1230

\bibitem[{{Wetterich}(1995)}]{Wetterich_1995}
{Wetterich}, C. 1995, \aap, 301, 321

\bibitem[{{Zhao} {et~al.}(2023){Zhao}, {Liu}, {Liao}, {Zhang}, {Liu}, \& {Du}}]{Zhao2023}
{Zhao}, Y., {Liu}, Y., {Liao}, S., {et~al.} 2023, \mnras, 523, 5962

\end{thebibliography}

\end{document}